\documentstyle[12pt]{article}
\newcommand{\ltapprox}{\raisebox{-0.6ex}
                       {$\ \textstyle \stackrel{\textstyle <}{\sim}\ $}
                      }

\def\figcap{\section*{Figure Captions\markboth
     {FIGURECAPTIONS}{FIGURECAPTIONS}}\list
     {Fig.
\arabic{enumi}:\hfill}{\settowidth\labelwidth{Fig. 999:}
     \leftmargin\labelwidth
     \advance\leftmargin\labelsep\usecounter{enumi}}}
 \relax
\def\reflist{\section*{REFERENCES\markboth
     {REFLIST}{REFLIST}}\list
     {[\arabic{enumi}]\hfill}{\settowidth\labelwidth{[999]}
     \leftmargin\labelwidth
     \advance\leftmargin\labelsep\usecounter{enumi}}}
 \relax
\def\tabcap{\section*{Tables\markboth
     {TABLES}{TABLES}}\list
     {Table
\arabic{enumi}:\hfill}{\settowidth\labelwidth{Table 999:}
     \leftmargin\labelwidth
     \advance\leftmargin\labelsep\usecounter{enumi}}}
 \relax
\renewcommand{\thefootnote}{\fnsymbol{footnote}}
\begin{document} \begin{titlepage}
%---------------------------------------------------------%
\rightline{NTUTH-96-09}
\rightline{June 1996}
 \null
 \vskip 0.5in
\begin{center}
 \vspace{.5in}
  {\Large
    $Zb\bar{b}$ Loop  Correction with Charge 2/3 Singlet Quarks}
  \par
 \vskip 0.5in
\setcounter{footnote}{1}
 {\large
  \begin{tabular}[t]{c}
       Wei-Shu Hou \\
       Department of Physics, National Taiwan University \\
       Taipei, Taiwan, R.O.C. \\
    \\
       Hsien-chung Kao \\
       Institute of Physics, Academia Sinica \\
       Nankang, Taipei, Taiwan, R.O.C. \\
       and \\
       Department of Physics, Tamkang University\footnotemark \\
       Tan-Shui, Taiwan, R.O.C. \\
  \end{tabular}}

\footnotetext{Permanent address.}
 \par \vskip 1.0em
 \vspace{0.8cm}
 {\large\bf Abstract}
\end{center}
\quotation
 
We calculate the non-universal correction to the $Zb\bar{b}$ vertex in a 
simple extension of the Standard Model, where a charge $+2/3$ isosinglet 
quark is added to the standard spectrum.
Comparison is made with other solutions to $R_b$ (and $R_c$) that
demand particles lighter than $M_W$.
 
\endquotation
%
%\vspace{.2in}
% 
% 
%\vfill
%\mbox{}
%\footnotetext{Permanent address after Sept. 1, 1992:
%                     Department of Physics, National Taiwan University,\\
%     \hspace*{0.65cm} Taipei, Taiwan 10764, R.O.C.}
\end{titlepage}
 
\baselineskip=21truept plus 0.2truept minus 0.2truept
\renewcommand{\thefootnote}{\arabic{footnote}}
\setcounter{footnote}{0}
\pagestyle{plain}
\pagenumbering{arabic}
 
Current experimental data on
$R_b\ (\equiv \Gamma(Z\to b\bar b)/\Gamma(Z\to \mbox{hadrons}))$
and $R_c$ deviate significantly from Standard Model (SM) predictions [\ref{LEPEWWG}].  
It has been suggested that both discrepancies can be accounted 
for if one makes a minimal extension of the SM by adding an extra up-type 
isosinglet quark $Q$ [\ref{RbRc}].  
In this scenario, $R_c$
is reduced because of $c$-$Q$ mixing,
while $R_b$ is enhanced by a more elaborate mechanism.
The top quark is light, and remains hidden below $M_W$ (i.e. $m_t < M_W$)
by some fast decay mechanism, while the 180 GeV quark observed at the Tevatron
is identified as the dominantly singlet quark.
As a result, the effective top quark mass, defined as the equivalent $m_t$ 
that appears in the $Zb\bar{b}$ vertex correction within the Standard Model,
is reduced and thus $R_b$ is increased.
The main purpose of this paper is to perform a
detailed calculation of the non-universal correction
to the $Zb\bar{b}$ vertex when $Q$ is present. 
The novel feature is the presence of tree level flavor changing neutral current (FCNC)
$Z$-$t$-$Q$ couplings in the loop.

Adding an up-type isosinglet quark gives rise to new gauge invariant mass 
terms of the type  $M\, \bar{Q}^0_L Q^0_R$ and $M'_j\, \bar{Q}^0_L u^0_{jR}$, 
where $u^0_j$ denotes the gauge eigenstates of standard up-type quarks.  
Moreover, there are also mass terms of the form $m'_i\, \bar{u}^0_{iL} Q^0_R$
which arise from additional Yukawa couplings.
Since we do not expect the first generation to play
an important role in the present context,
we will ignore them throughout the paper.
We therefore have the following mass matrix 
for the up-type quarks [\ref{HH}],
\begin{equation}
{\bf M_0} =       \left( \begin{array}{cc}
                    m & m' \\ 
                   M' & M
          \end{array} \right),
\end{equation}
%where $m$ is a $2\times 2$ Yukawa mass matrix. ${\bf M_0}$
which can be diagonalized by a biunitary transformation as usual,
\begin{equation}
{\bf M} \equiv {\rm diag}(m_c,\ m_t,\ m_Q) = S_u^\dagger {\bf M_0} T_u.
\end{equation}

In terms of the mass eigenstates $u$ and $d$, the charged current takes
the form [\ref{Silva}],
\begin{equation}
\bar u_{\alpha L} \left(V_{\alpha j} \right) \gamma_\mu d_{jL},
\end{equation}
where $\alpha $ ranges over $c,\ t,\ Q$,
and $j = s,\ b$.
The quark mixing matrix $V$ is now $3\times 2$ and hence non-unitary.
Since $\vert V_{cb}\vert^2 \simeq (0.04)^2 \ll 1$,
we shall set $V_{cb}=0$ throughout the paper,
which leads to considerable simplification of our results.

The neutral current for $u$-type quarks now becomes [\ref{Silva}],
\begin{equation}
\bar u_{L} 
\left(U\ t_3^u - I\ s^2_W \right)
 \gamma_\mu u_{L} 
+ \bar u_{R}
\left(- I\ s^2_W  \right)
 \gamma_\mu u_{R},
\label{NC}
\end{equation}
where we have suppressed flavor indices, $U = V V^\dagger$,
and $s_W \equiv \sin\bar{\theta}_W$ is
the effective electroweak mixing angle at the $Z$-pole.
Only the isospin part is modified since only $Q^0_L$
is nonstandard.
Because $V$ is non-unitary, $U$ has off-diagonal terms.
We also note that in general, $\vert U_{cc}\vert < 1$, and can be used to
account for the apparent suppression of $R_c$ [\ref{RbRc}].

To calculate the $Zb\bar{b}$ vertex correction, we must evaluate the ten 
Feynman diagrams listed in Fig. 1.
We make use of the 
REDUCE code set up by Hou and Stuart [\ref{HS}],
which is in turn based on the package LERG-I [\ref{LERGI}].
The results are expressed in terms of two scalar integrals,
\begin{eqnarray}
B_0(p^2; m_1^2, m_2^2)
& = & \, \int {d^n k \over i \pi^2}
{1 \over \bigl(k^2 + m_1^2 \bigr) \left[ (k+p)^2 + m_2^2 \right]}, \nonumber\\
C_0(p_1^2, p_2^2, p_5^2; m_1^2, m_2^2, m_3^2)
& = & \, \int {d^nk \over i \pi^2}
{1 \over \bigl(k^2 + m_1^2 \bigr) \left[ (k+p_1)^2 + m_2^2 \right] 
\left[ (k+p_1+p_2)^2 + m_3^2 \right] }. \nonumber
\end{eqnarray}
The original program was
set up for a general study of FCNC
decays of the possible fourth generation $b'$ quark.
It was used by Lynn and Stuart  [\ref{LS}]
to calculate the $Zb\bar{b}$ vertex 
correction in the context of SM. 
They find that the genuine FCNC vertex diagrams are
well-behaved in the limit that the masses of
the two external quarks become equal,
and the resulting expression can be used without further ado.
On the other hand, diagrams containing fermion self-energies
individually would have spurious divergences in this limit,
since one has a propagator factor of $1/(m_b^2 - m_{b'}^2)$.
However, the sum of the four diagrams has a well defined limit,
and its value can be determined using the L'Hospital rule.

In the on-shell scheme, the UV divergence is taken care of by the counterterm [\ref{LS}],
\begin{equation}
(-2i g^3\cos\bar{\theta}_W t_3^d/16 \pi^2) B_0(0; M^2_W, M^2_W).
\label{CT}
\end{equation}
One could equivalently make a subtraction at $q^2 = M_Z^2$.
We shall not go into the details of the renormalization program,
since we are concerned with only the
internal quark mass dependent contribution.
However, it is easy to understand the origin of
the counterterm of eq. (\ref{CT}).
In ref. [\ref{HS}], it was checked in great detail using elementary 
Ward identities that all the divergent pieces cancel
except for a left-over piece coming from diagram $(c)$,
which takes the form
$(2i g^3\cos\bar{\theta}_W t_3^d/16 \pi^2) B_0(q^2; M^2_W, M^2_W)$.
In the FCNC case, the term is removed by the GIM mechanism,
but in the flavor diagonal case such as $Zb\bar b$ vertex,
subtraction and renormalization is needed.
Thus the counterterm in eq.~(\ref{CT}) is the most natural
one.

For diagrams (c)--(j) there is only one quark in the loop,
either $t$ or $Q$, and the flavor factors are $|V_{tb}|^2$ and $|V_{Qb}|^2$, respectively.
One therefore finds the contribution
\begin{equation}
  |V_{tb}|^2\ {F_L}^{(n)}(m_t) 
               + |V_{Qb}|^2\ {F_L}^{(n)}(m_Q),
\quad \quad \quad n = c,\ d,\ \ldots,\ j,
\end{equation}
where ${F_L}^{(n)}(m)$ is the SM result for diagram $n$,
in the notation of ref. [\ref{HS}].
On the other hand, as can be seen from eq. (\ref{NC}), 
the two internal quarks in diagrams $(a)$ and $(b)$ can be different,
which is an interesting feature of introducing isosinglet quarks.
For identical internal quarks, the flavor factors are
$V_{tb}^*  U_{tt}  V_{tb}$ and $V_{Qb}^*  U_{QQ}  V_{Qb}$,
while if the two internal quarks are different, 
the flavor factors are
$V_{tb}^*  U_{tQ}  V_{Qb}$ and $V_{Qb}^*  U_{Qt}  V_{tb}$.
 
It is useful to establish some relations between these flavor factors.
Using the fact that $U= V V^\dagger$ and $V^\dagger V = I$,
one finds that
$U V  =  V$.
%$V^\dagger U  =  V^\dagger$.
Since we take $V_{cb} =0$, this translates to
\begin{eqnarray}
&  U_{tt} V_{tb} \; & \;\, = \; 
V_{tb} - U_{tQ} V_{Qb}, \\
& U_{QQ}  V_{Qb} & \;\, = \; 
V_{Qb} -  U_{Qt} V_{tb}, %\\
\end{eqnarray}
in component form.
Multiplying by $V_{tb}^*$ and $V_{Qb}^*$, respectively,
from the hermitian nature of $U$, one finds that
$V_{Qb}^*  U_{Qt}  V_{tb}$ and $V_{tb}^* U_{tQ} V_{Qb}$ are
not only real, but equal to each other, hence, 
\begin{eqnarray}
& V_{tb}^*  U_{tt}  V_{tb} \; & \;\, = \; 
|V_{tb}|^2 - V_{tb}^* U_{tQ} V_{Qb}, \label{decomp} \\
& V_{Qb}^*  U_{QQ}  V_{Qb} & \;\, = \; 
|V_{Qb}|^2 - V_{tb}^* U_{tQ} V_{Qb}. %\nonumber 
\end{eqnarray}
We therefore have the following contribution from diagrams $(a)$ and $(b)$,
\begin{eqnarray}
&  & |V_{tb}|^2\ {F_L}^{(n)}(m_t) 
+ |V_{Qb}|^2\ {F_L}^{(n)}(m_Q)  \nonumber \\
& - & V_{tb}^* U_{tQ} V_{Qb}\,
\left\{F_L^{(n)}(m_t,m_t) - 2 F_L^{(n)}(m_t,m_Q) + F_L^{(n)}(m_Q,m_Q)
\right\}\Bigg\vert_{t_3^u}.
\end{eqnarray}
According to eq. (\ref{NC}), only the  $t_3^u$ part should be used for the calculation
of the last term.

Putting everything together,
we finally obtain
\begin{equation}
 F_L  =  |V_{tb}|^2\, {F_L}^{SM}(m_t) + |V_{Qb}|^2\, {F_L}^{SM}(m_Q) 
  - V_{tb}^* U_{tQ} V_{Qb}\, \sum_{n = a}^b \Delta F_L^{(n)}(m_t,m_Q),
\label{final}
\end{equation}
where 
\begin{equation}
\Delta F_L^{(n)}(m_t,m_Q) = \left\{
F_L^{(n)}(m_t,m_t) - 2 F_L^{(n)}(m_t,m_Q) + F_L^{(n)}(m_Q,m_Q)
                                                                \right\}\Bigg\vert_{t_3^u}.
\label{delta}
\end{equation}
Since $V$ is part of a $3\times 3$ unitary matrix, it can be parameterized as
\begin{equation}
V =       \left( \begin{array}{cc}
                    C_2 & 0 \\ -S_2S_3 & C_3 e^{i\delta}
                  \\ +S_2C_3 & S_3 e^{i\delta}
          \end{array} \right),
\end{equation}
where $S_i \equiv \sin\theta_i$, $C_i \equiv \cos\theta_i$,
and $S_2$, $S_3$ are the $c$-$Q$, $t$-$Q$ mixing angles, respectively.
Consequently, we have $|V_{tb}|^2 = C_3^2$, $|V_{Qb}|^2 = S_3^2$, and
$V_{tb}^* U_{tQ} V_{Qb} = C_2^2 C_3^2 S_3^2.$
Note that the phase $\delta$ is removable
in any case because of setting $V_{cb} = 0$.

Inspecting eq. (\ref{final}) , we see that the first two terms are identical to the case of
adding a fourth generation, with $Q = t'$ and $S_3 = \vert V_{t'b}\vert$.
${F_L}^{SM}$ is nothing but the full SM result of ref. [\ref{LS}],
expressed in terms of three ``universal functions",
\begin{eqnarray}
{F_L}^{SM} = {g^3 \over 32\pi^2 c_W}
\left\{ \left(t^u_3 - Q^u s_W^2\right) \rho - 2 t^d_3 c_W^2\, \Lambda
-2 t^d_3\, \Xi \right\}.
\end{eqnarray}
The last term appears only for $Zb\bar{b}$ vertex 
but not in $\gamma b\bar{b}$.
We note that the expressions given in ref. [\ref{LS}] contain three typos,
two of which can be identified simply by dimensionality.
For $m$ not much smaller than $M_W$,
the leading effect is ${F_L}^{SM}(m) \propto m^2/M_W^2$.

The $\Delta F_L$ term is specific to adding 
singlet quarks, and arises from diagrams $(a)$ and $(b)$ only.
We find that $\Delta F_L^{(n)}$ contributes only to the $\Xi$ term.
This is hardly surprising since adding the singlet quark $Q$ should 
affect only the $Zb\bar{b}$ vertex.
The explicit result is
\begin{eqnarray}
\Delta \Xi^{(a)} 
&= & - \frac{1}{q^2}\left(M_W^2 - q^2 - m_Q^2 \right)^2
C_0(q^2, 0, 0, m_Q^2 ,m_Q^2, M_W^2 )\nonumber\\
&\;& + \frac{2}{q^2}\left(M_W^2 - q^2 - m_Q^2 \right)
\left(M_W^2 - q^2 - m_t^2 \right)
C_0(q^2, 0, 0, m_Q^2 ,m_t^2, M_W^2 )\nonumber\\
&\;& - \frac{1}{q^2}\left(M_W^2 - q^2 - m_t^2 \right)^2
C_0(q^2, 0, 0, m_t^2 ,m_t^2, M_W^2 )\nonumber\\
&\;& + \frac{1}{2q^2}\left(2M_W^2 - 3q^2 - 2m_Q^2 \right)
\left[ B_0(q^2 ,m_Q^2 ,m_Q^2 ) -  B_0(q^2 ,m_Q^2 ,m_t^2 ) \right] \nonumber\\
&\;& + \frac{1}{2q^2}\left(2M_W^2 - 3q^2 - 2m_t^2 \right)
\left[ B_0(q^2 ,m_t^2 ,m_t^2 ) -  B_0(q^2 ,m_Q^2 ,m_t^2 ) \right], \nonumber\\
\Delta \Xi^{(b)}
& = &  \frac{m_Q^4}{2M_W^2}C_0(q^2, 0, 0, m_Q^2 ,m_Q^2, M_W^2 )
     - \frac{m_Q^2 m_t^2}{M_W^2}C_0(q^2, 0, 0, m_Q^2 ,m_t^2, M_W^2 )\nonumber\\
&\;& + \frac{m_t^4}{2M_W^2}C_0(q^2, 0, 0, m_t^2 ,m_t^2, M_W^2 ). \nonumber
\end{eqnarray}
The leading effect comes from $\Delta \Xi^{(b)}$, which takes on a rather 
simple form.  This can be easily understood as follows.  From diagram $(b)$ 
with only $t_3^u$ part of the $Z$-$\alpha$-$\beta$ vertex, 
\begin{eqnarray}
\Gamma_\mu^{(b)}(m_\alpha,m_\beta) 
& \propto & \int {d^4k \over (2\pi)^4}\;
{{\textstyle m_\alpha \over M_W} \gamma_R\; 
(\not{\! p'} + \not{\! k} + m_\alpha)\;
t^u_3 \gamma_\mu \gamma_L\;
(\not{\! p} + \not{\! k} + m_\beta)\;
{\textstyle m_\beta \over M_W} \gamma_L
 \over \left(k^2 + M_W^2 \right) \left[(k+p')^2 + m_\alpha^2 \right] 
\left[ (k+p)^2 + m_\beta^2 \right] } \nonumber \\
      & = & {m_\alpha^2 m_\beta^2\over M_W^2}\,
\int {d^4k \over (2\pi)^4}\;
{t^u_3 \gamma_\mu \gamma_L 
\over \left(k^2 + M_W^2 \right) \left[(k+p')^2 + m_\alpha^2 \right] 
\left[ (k+p)^2 + m_\beta^2 \right] }  \nonumber\\
& \propto &  {m_\alpha^2 m_\beta^2\over M_W^2}\;
C_0(0, q^2, 0; M_W^2, m_\alpha^2, m_\beta^2 ).
\end{eqnarray}
The sign of  the $\Delta F_L$ term can be understood as follows.
As noted earlier, the first two terms of 
eq. (\ref{final}) are identical to the case of adding a fourth generation,
hence the $\Delta F_L$ term represents the difference between singlet and 
fourth generation results.
As $F_L$ is mainly a measure of nondecoupling,
one expects the $\Delta F_L$ term to soften the effect, since in the 
singlet case the gauge invariant masses $M'$ and $M$
of eq. (1) are of decoupling nature,
in contrast to the Yukawa masses $m$ and $m'$.
Note that $\Delta F_L$ has a built-in cancelation mechanism,
hence it is unlikely to be dominant.
In Fig. 2 we compare $F_L^{SM}(m_t)$ and $\Delta F_L(m_t,m_Q)$
vs. $m_t$ for $m_Q$ fixed at 180 GeV.  Since $S_3^2$ should not 
be too large, the $\Delta F_L$ term is indeed subdominant.  

The non-universal correction to the $Zb\bar{b}$ width is [\ref{db}]
\begin{equation}
\Gamma(Z \to b\bar{b}) = \Gamma^0_b \left[1 + \delta_b \right],
\end{equation}
where $\Gamma^0_b$ includes all effects
other than the one specific to the $Zb\bar{b}$ vertex.
As a first approximation, taking leading effect only, our result  is equivalent to an effective top mass 
\begin{equation}
(m_t^{eff.})^2 \approx C_3^2 m_t^2 + S_3^2 m_Q^2 = m_t^2 + S_3^2(m_Q^2 - m_t^2),
\label{mtef}
\end{equation}
in the SM $Zb\bar b$ loop, i.e. $\delta_b^{SM}(m_t^{eff.})$,
which can be fitted by [\ref{db}]
\begin{equation}
\delta_b^{SM}(m_t^{eff.}) = 0.01\left( -0.49 {(m^{eff}_t)^2 \over M_Z^2} - 0.45 \right),
\label{dbsm}
\end{equation}
in the mass range of interest.
The constant term differs from that of ref. [\ref{db}]
since it  is scheme dependent.
In Fig. 3 we compare our full result and the approximate
one of eq. (\ref{dbsm}). % in terms of $\delta_b$ vs. $S_3^2$.
The full result is smaller in absolute value by about 6 to 14 percent
for given $S_3^2$ value, in agreement with qualitative arguments given earlier.

Since eq. (\ref{mtef}) is different from what was used in 
ref. [\ref{RbRc}], it is worthwhile to repeat the numerical estimates.
$S_2^2$ is not affected and has the value $0.03$ from taking the $R_c$ value of 0.161.
To fit $R_b \cong 0.2219$  with $m_t,\ m_Q =70,\ 180$ GeV,
but using eqs. (\ref{mtef}) and (\ref{dbsm}), we find that
$m_t^{eff.} \simeq 95$ GeV and $S_3^2 \simeq 0.15$.
The numerical value of $S_3^2$, however, depends quite sensitively on $m_t^{eff}$,
and hence on the small subleading corrections that we have computed.
Using the full result, we find from inspection of Fig. 3 that
the actual $S_3^2$ is around 0.22.
These numbers are not drastically different from ref. [\ref{RbRc}].

It is argued in ref. [\ref{RbRc}] that in order to hide the $t$ quark and 
at same time keep $R_l$ untouched, one has to introduce an additional Higgs doublet.
As a result, there are additional corrections to the $Zb\bar{b}$ vertex 
from diagrams $(b)$,\ $(d)$,\ $(i)$ and $(j)$ due to the physical charged Higgs boson.
Since the mass of the physical charged Higgs is 
taken to be much greater [\ref{RbRc}] than $M_W$,
we expect its correction to the $Zb\bar{b}$ vertex
to be suppressed.
Indeed, by taking the physical charged Higgs mass to be 250 GeV,
we have checked numerically that it only adds about 6-11 percent to
the previous result of  $\delta_b$, since $\cot\beta$ is constrained by $B$-mixing.

We turn to some comments and discussion.
The scenario of ref. [\ref{RbRc}] relates the $R_b$ and $R_c$ problem
to the existence of heavy singlet quark $Q$ and light top quark.
It is similar to the partial SUSY solution to $R_b$ (but not $R_c$)
in predicting a host of scalars and fermions below $M_W$.
It is, however,  in spirit closer to fourth generation models  [\ref{GMP},\ref{CHW}].
Note that for the fourth generation case,
assuming that $\vert V_{tb}\vert^2 + \vert V_{t'b}\vert^2 = 1$, one has
\begin{eqnarray}
 F_L = {F_L}^{SM}(m_t)
 + \vert V_{t^\prime b}\vert^2\ [{F_L}^{SM}(m_{t^\prime}) - {F_L}^{SM}(m_t)].
\end{eqnarray}
Since the absolute value of ${F_L}^{SM}(m)$ increases as $m^2$,
taking $m_t \sim 180$ GeV and $m_{t^\prime} > m_t$
would aggravate the $R_b$ problem.
To alleviate or resolve the $R_b$ problem, one must have
either $m_t = 180$ GeV and $m_{t^\prime} < m_t$  [\ref{GMP}],
or $m_{t^\prime} = 180$ GeV and  $m_t < m_{t^\prime}$  [\ref{CHW}],
in the convention that $t$ is the isospin partner of the $b$ quark.
The models of refs.  [\ref{GMP}] and [\ref{CHW}] also contain light supersymmetric particles,
such as  stop $\tilde t$ and chargino/neutralino $\chi^\pm$, $\chi^0$,
which are needed either to hide the lighter quark,
or explain the $R_b$ problem.

Thus, in all these cases, light particles are predicted to exist below $M_W$,
and should be readily discovered as LEP 1.6 turns on and accumulates 
sufficient integrated luminosity.
As we eagerly await imminent discovery, we should keep in mind
that all these scenarios could be ruled out before the end of 1996.
The implication would be that the $R_b$ problem cannot be explained
in the context of these models.
It should be stressed, however, that the singlet model stands out
in its ability to suppress $R_c$ with relative ease,
without necessarily touching $R_b$.
Experimentally, $R_c$ is harder to measure,
hence it would take a long time before one can be confident that
the experimental value is in full agreement with SM.
Assuming that no light particles are found at LEP 1.6 and beyond,
we could take $m_Q > m_t = 180$ GeV and fit any small deviation in $R_c$.
If we take $S_3^2 \sim S_2^2 \ltapprox 0.03$,
and/or if $m_Q \sim m_t$, we see from
eq. (\ref{final}) that $R_b$ is not much affected.
It is amusing that if $m_Q$ is not very different from $m_t$,
it could partially explain the larger ``$\sigma_{t\bar t}$" observed by CDF [\ref{Kruse}].
We urge the LEP experiments to continue the refinement of $R_c$
(and of course $R_b$) measurement.

In this letter, we present a calculation of the non-universal correction 
to the $Zb\bar{b}$ vertex when a charge $+2/3$ isosinglet quark is added to 
the Standard Model.  Since the GIM mechanism is violated, it is possible 
to have flavor changing neutral current $Z$-$t$-$Q$ couplings in the loop.
The result is close to but slightly weaker than that of adding
a fourth generation, and can be approximated as an effective top mass.
If we identify the dominantly singlet quark to be the one observed at the Tevatron 
and assume the top quark to be lighter than $M_W$,
we can fit the the current experimental
data of $R_b$ (and $R_c$) by choosing appropriate $S_3^2$ (and $S_2^2$) values.
If LEP 1.6 does not find the light top quark, 
the model cannot explain $R_b$ but could still account for
suppression of $R_c$, even if $m_Q > 180$ GeV.

\vskip 0.5in
\noindent{\bf Acknowledgments}
WSH is supported in part by grant NSC 85-2112-M-002-011 and HCK by
NSC 85-2112-M-001-004 of the Republic of China. 

\vfill\eject

\begin{reflist}
\item \label{LEPEWWG} LEP Electroweak Working Group Report, LEPEWWG/96-01 (1996).
\item \label{RbRc} G. Bhattacharyya, G. C. Branco and W. S. Hou,
hep-ph/9512239, to appear in Phys. Rev. D.
\item \label{HH} W. S. Hou and H. C. Huang,  Phys. Rev. D {\bf 51}, 5285
(1995).
\item \label{Silva} L. Lavoura and J. P. Silva,  Phys. Rev. D {\bf 47}, 
2046 (1993).
\item \label{HS} W. S. Hou and R. G. Stuart, Nucl. Phys. {\bf B} 320 (1989) 277.
\item \label{LERGI} R. G. Stuart, Comput. Phys. Commun. 48 (1988) 569 .
\item \label{LS} B. W. Lynn and R. G. Stuart, Phys. Lett. {\bf B} 252 (1990) 676.
\item \label{db} J. Bernab\'eu, A. Pich and A. Santamaria, Nucl. Phys. {\bf B} 363 (1996) 326. 
\item \label{GMP} J. F. Gunion, D. W. McKay and H. Pois, Phys. Rev. D {\bf 53} 1616 (1996). 
\item \label{CHW} M. Carena, H. E. Haber and C. E. M. Wagner, hep-ph/9512446. 
\item \label{Kruse} M. Kruse, talk given at
{\it XI Topical Workshop on Proton-Antiproton Collider Physics},
Abano Terme, Italy, May 1996.
\end{reflist}

\vfill\eject
    
\begin{figure}[h]
\includegraphics{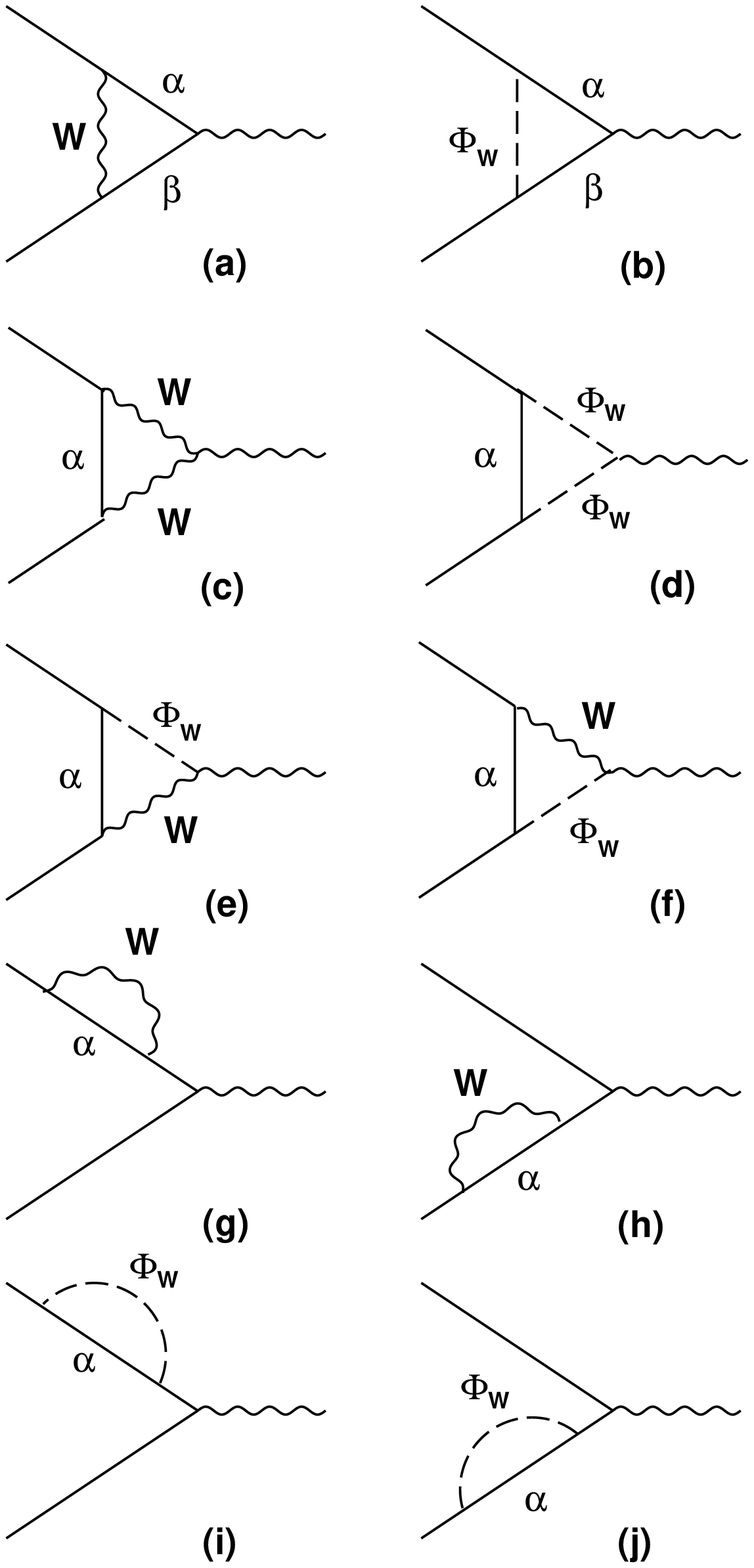}
\vskip 16.5cm
\end{figure}
$\qquad $ 
Fig.~1  The ten Feynman diagrams that contribute to the $Zb\bar{b}$ vertex.

\vfill\eject

\begin{figure}[h]
\includegraphics{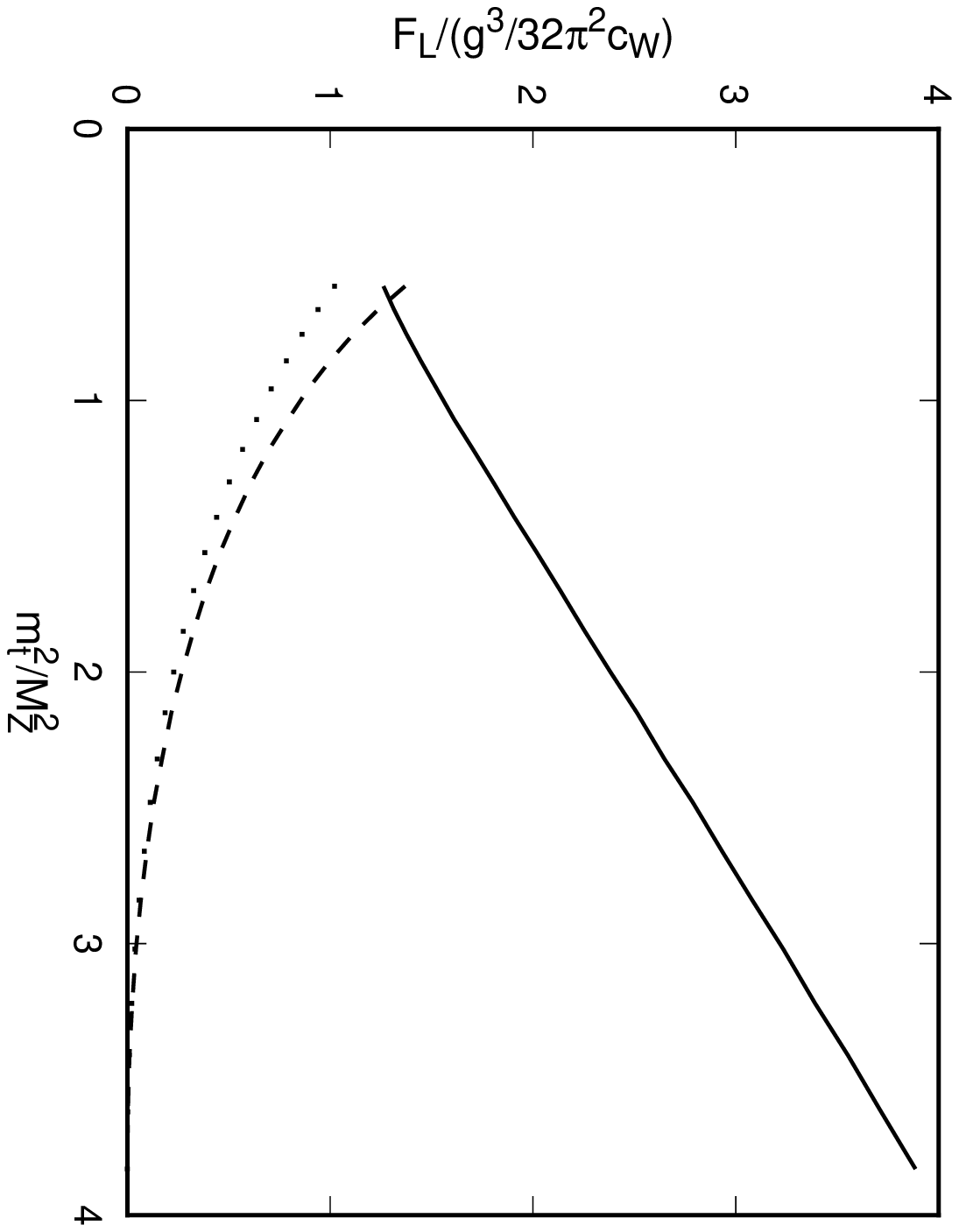}
\vskip 12.5cm
\end{figure}
Fig.~2  Comparison of $F_L^{SM}(m_t)$ (solid) and 
$\Delta F_L^{(b)}(m_t,m_Q)$, $ \sum_{n = a}^b \Delta F_L^{(n)}(m_t,m_Q)$,
(dots, dash) vs. $m_t$ for $m_Q = 180$ GeV.

\vfill\eject

\begin{figure}[h]
\includegraphics{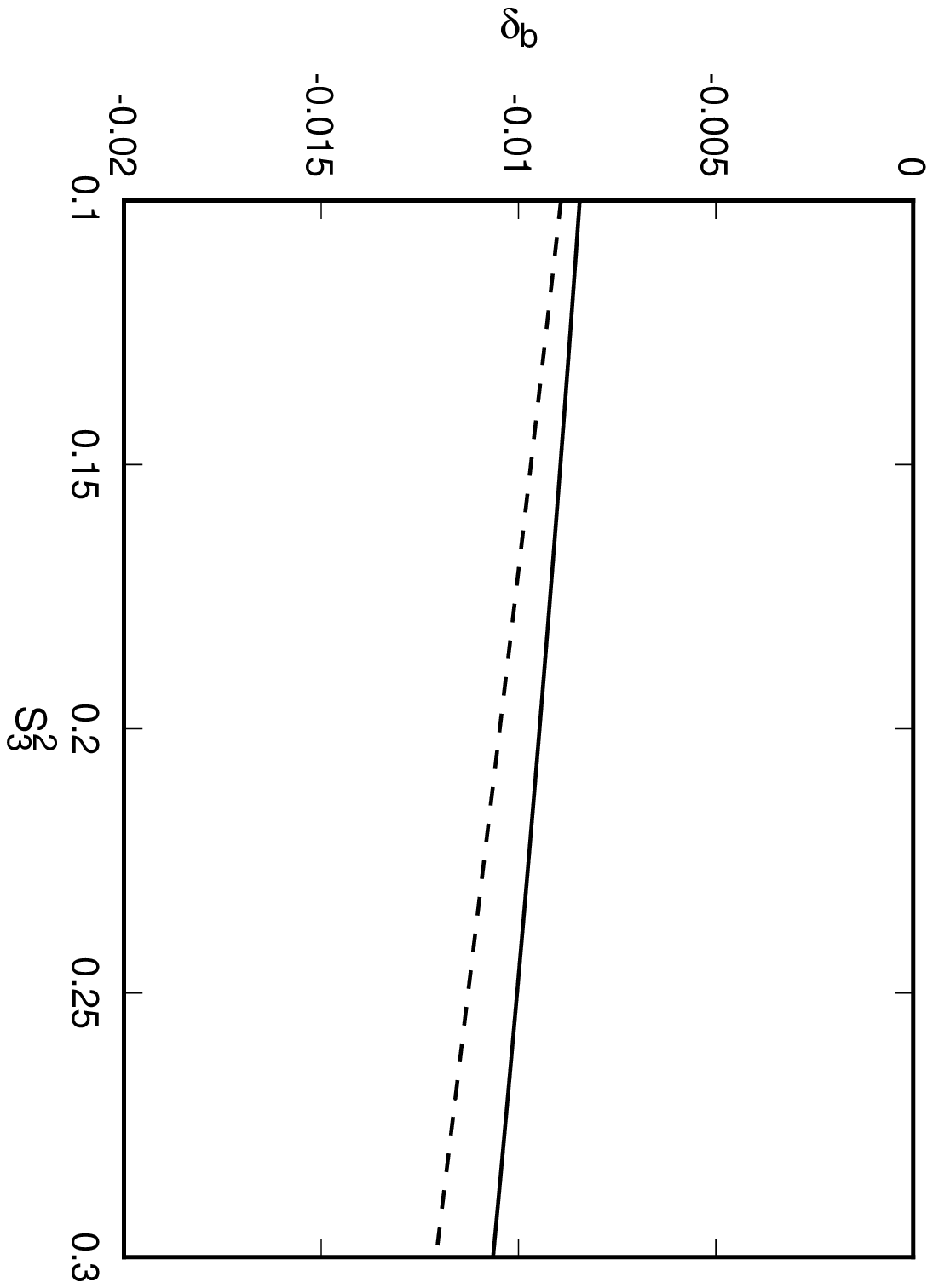}
\vskip 12.5cm
\end{figure}
Fig.~3  Comparison between the full result (solid) and the approximate one
(dash).  
$S_2^2$ is fixed at 0.03, while $S_3^2$ varies between 0.1 and 0.3,
with $m_t$ and $m_Q$ fixed at 70, 180 GeV, respectively.

\end{document}